\documentclass{sig-alternate}
\usepackage{times}
\usepackage{color}
\usepackage{url}
\usepackage{graphicx}
\usepackage{subfigure}
\usepackage[boxed,vlined]{algorithm2e}

\begin{document}

\newtheorem{theorem}{Theorem}
\newtheorem{corollary}{Corollary}
\newtheorem{fassumption}{Fundamental Assumption}
\newtheorem{lemma}{Lemma}
\newtheorem{algo}{Algorithm}

\newcommand{\hide}[1]{}
\newcommand{\semihide}[1]{{\tiny #1}}
\newcommand{\rev}[1]{{{\textcolor{red}{[[#1]]}}}}
\newcommand{\xhdr}[1]{\vspace{1.7mm}\noindent{{\bf #1.}}}

\newcommand{\denselist}{ \itemsep -5pt\topsep-10pt\partopsep-10pt }
\newcommand{\denselistA}{ \itemsep -0.8pt\topsep-5pt\partopsep-7pt }

\def\s{{s}}    
\def\D{{D}}    
\def\d{{d}}    
\def\L{{L}}    
\def\l{{l}}    
\def\w{{a}}    
\newcommand{\wij}[1]{{a}_{{#1}}}
\def\Q{{Q}}    
\def\dummy{{q}}   
\def\atr{{\psi}}  
\newcommand{\attrij}[1]{{\psi}_{{#1}}}
\newcommand{\atrij}[1]{{\psi}_{{#1}}}
\def\p{{p}}    
\def\prm{{w}}  
\def\b{{b}}    
\def\sk{{\xi}} 
\newcommand{\deriv}[2]{\frac{\partial #2}{\partial #1}} 

\conferenceinfo{WSDM'11,} {February 9--12, 2011, Hong Kong, China.}
\CopyrightYear{2011}
\crdata{978-1-4503-0493-1/11/02}
\clubpenalty=10000
\widowpenalty = 10000

\title{
Supervised Random Walks:\\Predicting and Recommending Links in Social Networks
}

\numberofauthors{2}
\author{
\alignauthor
Lars Backstrom\\
\affaddr{Facebook}\\
\email{lars@facebook.com}
\alignauthor
Jure Leskovec\\
\affaddr{Stanford University}\\
\email{jure@cs.stanford.edu}
}

\maketitle

\begin{abstract}
Predicting the occurrence of links is a fundamental problem in networks. In the
link prediction problem we are given a snapshot of a network and would like to
infer which interactions among existing members are likely to occur in the near
future or which existing interactions are we missing. Although this problem has
been extensively studied, the challenge of how to effectively combine the
information from the network structure with rich node and edge attribute data
remains largely open.

We develop an algorithm based on {\em Supervised Random Walks} that naturally
combines the information from the network structure with node and edge level
attributes. We achieve this by using these attributes to guide a random walk on
the graph. We formulate a supervised learning task where the goal is to learn a
function that assigns strengths to edges in the network such that a random
walker is more likely to visit the nodes to which new links will be created in
the future. We develop an efficient training algorithm to directly learn the
edge strength estimation function.

Our experiments on the Facebook social graph and large collaboration networks
show that our approach outperforms state-of-the-art unsupervised approaches as
well as approaches that are based on feature extraction.
\end{abstract}

\vspace{2mm} \noindent {\bf Categories and Subject Descriptors:} H.2.8 {\bf
[Database Management]}: Database applications---{\it Data mining}

\vspace{1mm} \noindent {\bf General Terms:} Algorithms; Experimentation.

\vspace{1mm} \noindent {\bf Keywords:} Link prediction, Social networks

\section{Introduction}
\label{sec:intro}
Large real-world networks exhibit a range of interesting properties and
patterns~\cite{barabasi99emergence,jure05dpl}. One of the recurring themes in
this line of research is to design models that predict and reproduce the
emergence of such network structures. Research then seeks to develop models
that will accurately predict the global structure of the
network~\cite{barabasi99emergence,jure05dpl,jure08microevol,backstrom06groups}.

Many types of networks and especially social networks are highly dynamic; they
grow and change quickly through the additions of new edges which signify the
appearance of new interactions between the nodes of the network. Thus, studying
the networks at a level of individual edge creations is also interesting and in
some respects more difficult than global network modeling. Identifying the
mechanisms by which such social networks evolve at the level of individual
edges is a fundamental question that is still not well understood, and it forms
the motivation for our work here.

We consider the classical problem of link
prediction~\cite{libennowell03linkpred} where we are given a snapshot of a
social network at time $t$, and we seek to accurately predict the edges that
will be added to the network during the interval from time $t$ to a given
future time $t'$. More concretely, we are given a large network, say Facebook,
at time $t$ and for each user we would like to predict what new edges
(friendships) that user will create between $t$ and some future time $t'$. The
problem can be also viewed as a {\em link recommendation} problem, where we aim
to suggest to each user a list of people that the user is likely to create new
connections to. 

The processes guiding link creation are of interest from more
than a purely scientific point of view.  The current Facebook system for
suggesting friends is responsible for a significant fraction of link creations,
and adds value for Facebook users.  By making better predictions, we will be
able to increase the usage of this feature, and make it more useful to Facebook
members.

\xhdr{Challenges} The link prediction and link recommendation problems are
challenging from at least two points of view. First, real networks are
extremely sparse, i.e., nodes have connections to only a very small fraction of
all nodes in the network. For example, in the case of Facebook a typical user
is connected to about 100 out of more than 500 million nodes of the network.
Thus, a very good (but unfortunately useless) way to predict edges is to
predict {\em no new edges} since this achieves near perfect predictive accuracy
(i.e., out of 500 million possible predictions it makes only 100 mistakes).

The second challenge is more subtle; to what extent can the links of the social
network be modeled using the features intrinsic to the network itself?
Similarly, how do characteristics of users (\emph{e.g.}, age, gender, home
town) interact with the creation of new edges? Consider the Facebook social
network, for example. There can be many reasons exogenous to the network for
two users to become connected: it could be that they met at a party, and then
connected on Facebook. However, since they met at a party they are likely to be
about the same age, and they also probably live in the same town. Moreover,
this link might also be hinted at by the structure of the network: two people
are more likely to meet at the same party if they are ``close'' in the network.
Such a pair of people likely has friends in common, and travel in similar
social circles. Thus, despite the fact that they became friends due to the
exogenous event (\emph{i.e.}, a party) there are clues in their social networks
which suggest a high probability of a future friendship.

Thus the question is how do network and node features interact in the creation
of new links. From the link creation point of view: how important is it to have
common interests and characteristics? Furthermore, how important is it to be in
the same social circle and be ``close'' in the network in order to eventually
connect. From the technical point of view it is not clear how to develop a
method that, in a principled way, combines the features of nodes (\emph{i.e.},
user profile information) and edges (\emph{i.e.}, interaction information) with
the network structure. A common, but somewhat unsatisfactory, approach is to
simply extract a set of features describing the network structure (like node
degree, number of common friends, shortest path length) around the two nodes of
interest and combine it with the user profile information.

\xhdr{Present work: Supervised Random Walks} To address these challenges we
develop a method for both link prediction and link recommendation. We develop a
concept of {\em Supervised Random Walks} that naturally and in a principled way
combines the network structure with the characteristics (attributes, features)
of nodes and edges of the network into a unified link prediction algorithm.

We develop a method based on {\em Supervised Random Walks} that in a {\em
supervised} way learns how to bias a PageRank-like random walk on the
network~\cite{agarwal06rank,agarwal07rank} so that it visits given nodes
(\emph{i.e.}, positive training examples) more often than the others.


We achieve this by using node and edge features to learn edge {\em strengths}
(\emph{i.e.}, random walk transition probabilities) such that the random walk
on a such weighted network is more likely to visit ``positive'' than
``negative'' nodes. In the context of link prediction, positive nodes are nodes
to which new edges will be created in the future, and negative are all other
nodes.
We formulate a supervised learning task where we are given a source node $s$
and training examples about which nodes $s$ will create links to in the future.
The goal is to then learn a function that assigns a strength (\emph{i.e.},
random walk transition probability) to each edge so that when computing the
random walk scores in such a weighted network nodes to which $s$ creates new
links have higher scores to $s$ than nodes to which $s$ does not create links.

From a technical perspective, we show that such edge strength function can be
learned directly and efficiently. This means, that we do not 
postulate what it means for edge to be ``strong'' in an ad-hoc way and then use this
heuristic estimate. Rather, we show how to directly find the parameters of the
edge strength function which give optimal performance. This means we are able
to compute the gradient of the parameters of the edge strength function with
respect to the PageRank-like random walk scores. The formulation results in an
optimization problem for which we derive an efficient estimation procedure.

From the practical point of view, we experiment with large collaboration
networks and data from the Facebook network, showing that our approach
outperforms state-of-the-art unsupervised approaches as well as supervised
approaches based on complex network feature extraction. An additional benefit
of our approach is that no complex network feature extraction or domain
expertise are necessary as our algorithm nicely combines the node attribute and
network structure information.



\xhdr{Applications and consequences} As networks evolve and grow by addition of
new edges, the link prediction problem offers insights into the factors behind
creation of individual edges as well as into network formation in general.

Moreover, the link-prediction and the link-recommendation problems are relevant
to a number of interesting current applications of social networks. First, for
online social networking websites, like Facebook and Myspace, being able to
predict future interactions has direct business consequences. More broadly,
large organizations can directly benefit from the interactions within the
informal social network among its members and link-prediction methods can be
used to suggest possible new collaborations and interactions within the
organization. Research in security has recently recognized the role of social
network analysis for this domain (e.g., terrorist networks). In this context link
prediction can be used to suggest the most likely links that may form in the
future. Similarly, link prediction can also be used for prediction of missing
or unobserved links in networks~\cite{clauset08hierarchical} or to suggest
which individuals may be working together even though their interaction has yet
been directly observed. Applications go well beyond social networks, as our
techniques can be used to predict unobserved links in protein-protein
interaction networks in systems biology or give suggestions to bloggers about
which relevant pages on the Web to link to.

Furthermore, the framework we develop is more general than link prediction, and
could be used for any sort of interaction. For instance, in a collaboration
network, it could easily be used not to predict who $s$ will link to next
(write a paper with a previously un-collaborated-with person) but to predict
who $s$ will coauthor a paper with next, including all those with whom $s$ has
previously coauthored.

\xhdr{Further related work} The link prediction problem in networks comes in
many flavors and variants. For example, the network inference
problem~\cite{gomez10netinf,myers10connie} can be cast as a link prediction
problem where no knowledge of the network is given. Moreover, even models of
complex networks, like Preferential Attachment~\cite{barabasi99emergence},
Forest Fire model~\cite{jure05dpl} and models based on random
walks~\cite{jure08microevol,blum06surfer}, can be viewed as ways for predicting
new links in networks.

The unsupervised methods for link prediction were extensively evaluated by
Liben-Nowell and Kleinberg~\cite{libennowell03linkpred} who found that the
Adamic-Adar measure of node similarity~\cite{adamic03} performed best. More
recently approaches based on network community
detection~\cite{clauset08hierarchical,henderson09link} have been tested on
small networks. Link prediction in supervised machine learning setting was
mainly studied by the relational learning
community~\cite{taskar03linkpred,popescul03linkpred}. However, the challenge
with these approaches is primarily scalability.

Random walks on graphs have been considered for computing node proximities in
large graphs~\cite{tong06fast,tong07proximity,tong06subgraphs,sarkar09fast}.
They have also been used for learning to rank nodes in
graphs~\cite{agarwal06rank,agarwal07rank,minkov07rank,diligenti2005learning}.


\hide{

The link-prediction problem is also related to the problem of inferring missing
links from an observed network: in a number of domains, one constructs a
network of interactions based on observable data and then tries to infer
additional links that, while not directly visible, are likely to exist [10, 32,
36]. This line of work dirs from our problem formulation in that it works with
a static snapshot of a network, rather than considering network evolution; it
also tends to take into account specic attributes of the nodes in the network,
rather than evaluating the power of prediction methods that are based purely on
the graph structure.

Our goal is to make this intuitive notion precise, and to understand which
measures of proximity in a network lead to the most accurate link predictions.

We find that a number of proximity measures lead to predictions that outperform
chance by factors of forty to fty, indicating that the network topology does
indeed contain latent information from which to infer future interactions.

Moreover, certain fairly subtle measures|involving innite sums over paths in
the network|often outperform more direct measures, such as shortest-path
distances and numbers of shared neighbors.

Second problem is more subtle and deals with the interaction of the
characteristic of nodes and edges of the network and the structure of the
network itself.

From the machine learning point of view link prediction is hard for several
reasons.

Reasoning:
\begin{itemize}
  \item Link prediction problem
    \item Hard to define precisely as links are coming one by one
    \item Liben-Nowell first defined the problem as a ranking problem but
        they only used simple node-degree based features to compute the
        most likely edges to be added next.
    \item The challenge then is to combine both the node information
        together with the network information. Extracting node and network
        features would be a possible approach but it is very heuristic and
        unsatisfactory.
    \item The second challenge is evaluation
    \item Hard to get network data where we have information about links
        that appear over time and also rich information
        (attributes/features) about individual nodes and edges.
    \item We propose to combine both the network structure and node/edge
        attribute data into a single model that we use for link prediction.
        The idea is to use Random Walks with Restarts~\cite{tong06fast} to
        rank nodes on a graph but we then also use node/edge attribute data
        to learn edge transition probabilities such that nodes to which
        links are created are more likely to be visited by a random walked
        than nodes to which edges are not created.
    \item Random walks have been found to be good models for social
        networks~\cite{jure08microevol,blum06surfer}.
\end{itemize}

} 

\section{
Supervised Random Walks}
\label{sec:proposed}
Next we describe our algorithm for link prediction and recommendation. The
general setting is that we are given a graph and a node $\s$ for which we would
like to predict/recommend new links. The idea is that $s$ has already created
some links and we would like to predict which links it will create next (or
will be created to it, since the direction of the links is often not clear).
For simplicity the following discussion will focus on a single node $\s$ and
how to predict the links it will create in the future.

Note that our setting is much more general than it appears. We require that for
a node $\s$ we are given a set of ``positive'' and ``negative'' training nodes
and our algorithm then learns how to distinguish them. This can be used for
link prediction (positive nodes are those to which links are created in the
future), link recommendation (positive nodes are those which user clicks on),
link anomaly detection (positive nodes are those to which $s$ has anomalous
links) or missing link prediction (positive nodes are those to which $s$ has
missing links), to name a few. Moreover, our approach can also be generalized
to a setting where prediction/recommendation is not being made for only a
single node $s$ but also for a group of nodes.

\xhdr{General considerations} A first general approach to link prediction would
be to view it as a classification task. We take pairs of nodes to which $\s$
has created edges as positive training examples, and all other nodes as
negative training examples. We then learn a classifier that predicts where node
$\s$ is going to create links. There are several problems with such an
approach. The first is the class imbalance; $\s$ will create edges to a very
small fraction of the total nodes in the network and learning is particularly
hard in domains with high class imbalance. Second, extracting the features that
the learning algorithm would use is a challenging and cumbersome task. Deciding
which node features (\emph{e.g.}, node demographics like, age, gender,
hometown) and edge features (\emph{e.g.}, interaction activity) to use is
already hard. However, it is even less clear how to extract good features that
describe the network structure and patterns of connectivity between the pair of
nodes under consideration.

Even in a simple undirected graph with no node/edge attributes, there are
countless ways to describe the proximity of two nodes.  For example, we might
start by counting the number of common neighbors between the two nodes.  We
might then adjust the proximity score based on the degrees of the two nodes
(with the intuition being that high-degree nodes are likely to have common
neighbors by mere happenstance).  We might go further giving different length
two paths different weights based on things like the centrality or degree of
the intermediate nodes.  The possibilities are endless, and extracting useful
features is typically done by trial and error rather than any principled
approach.
The problem becomes even harder when annotations are added to edges. For
instance, in many networks we know the creation times of edges, and this is
likely to be a useful feature.  But how do we combine the creation times of all
the edges to get a feature relevant to a pair of nodes?

A second general approach to the link prediction problem is to think about it
as a task to rank the nodes of the network. The idea is to design an algorithm
that will assign higher scores to nodes which $\s$ created links to than to
those that $\s$ did not link to. PageRank~\cite{page98pagerank} and variants
like Personalized PageRank~\cite{jeh03personalized,haveliwala02topic} and Random
Walks with Restarts~\cite{tong06fast} are popular methods for ranking
nodes on graphs. Thus, one simple idea would be to start a random walk at node
$\s$ and compute the proximity of each other node to node
$\s$~\cite{tong07proximity}. This can be done by setting the random jump vector
so that the walk only jumps back to $\s$ and thus restarts the walk. The
stationary distribution of such random walk assigns each node a score
(\emph{i.e.}, a PageRank score) which gives us a ranking of how ``close'' to
the node $\s$ are other nodes in the network. This method takes advantage of
the structure of the network but does not consider the impact of other
properties, like age, gender, and creation time.

\xhdr{Overview of our approach} We combine the two above approaches into a
single framework that will at the same time consider rich node and edge
features as well as the structure of the network. As Random Walks with Restarts
have proven to be a powerful tool for computing node proximities on graphs we
use them as a way to consider the network structure. However, we then use the
node and edge attribute data to bias the random walk so that it will more often
visit nodes to which $\s$ creates edges in the future.

More precisely, we are given a source node $\s$. Then we are also given a set
of {\em destination nodes} $\d_1, \dots, \d_k \in D$ to which $\s$ will create
edges in the near future. Now, we aim to bias the random walk originating from
$s$ so that it will visit nodes $\d_i$ more often than other nodes in the
network. One way to bias the random walk is to assign each edge a random walk
transition probability (\emph{i.e.}, strength). Whereas the traditional
PageRank assumes that transition probabilities of all edges to be the same, we
learn how to assign each edge a transition probability so that the random walk
is more likely to visit target nodes $\d_i$ than other nodes of the network.
However, directly setting an arbitrary transition probability to each edge
would make the task trivial, and would result in drastic overfitting. Thus, we
aim to learn a model (a function) that will assign the transition probability
for each edge $(u,v)$ based on features of nodes $u$ and $v$, as well as the
features of the edge $(u,v)$. The question we address next is, how to directly
and in a principled way estimate the parameters of such random walk biasing
function?

\xhdr{Problem formulation} We are given a directed graph $G(V,E)$, a node $\s$
and a set of candidates to which $\s$ could create an edge. We label nodes to
which $\s$ creates edges in the future as {\em destination nodes} $\D=\{\d_1,
\dots, \d_k\}$, while we call other nodes to which $\s$ does not create edges
{\em no-link nodes} $\L=\{\l_1, \dots, \l_n\}$. We label candidate nodes with a
set $C=\{c_i\}=\D \cup \L$. We think of nodes in $D$ as positive and nodes in
$L$ as negative training examples.
Later we generalize to multiple instances of $s$, $L$ and $D$. Each node and
each edge in $G$ is further described with a set of features.  We assume that each edge $(u,v)$ has a corresponding feature vector
$\atrij{uv}$ that describes the nodes $u$ and $v$ (\emph{e.g.}, age, gender,
hometown) and the interaction attributes (\emph{e.g.}, when the edge was
created, how many messages $u$ and $v$ exchanged, or how many photos they
appeared together in).

For edge $(u,v)$ in $G$ we compute the strength
$\wij{uv}=f_{\prm}(\atrij{uv})$. Function $f_{\prm}$ parameterized by $\prm$
takes the edge feature vector $\atrij{uv}$ as input and computes the
corresponding edge strength $\wij{uv}$ that models the random walk transition
probability. It is exactly the function $f_{\prm}(\atr)$ that we learn in the
training phase of the algorithm.

To predict new edges of node $s$, first edge strengths of all edges are
calculated using $f_{\prm}$. Then a random walk with restarts is run from $s$.
The stationary distribution $\p$ of the random walk assigns each node $u$ a
probability $\p_u$.  Nodes are ordered by $\p_u$ and top ranked nodes are then
predicted as destinations of future links of $s$.

Now our task is to learn the parameters $\prm$ of function
$f_{\prm}(\atrij{uv})$ that assigns each edge a transition probability
$\wij{uv}$. One can think of the weights $\wij{uv}$ as edge strengths and the
random walk is more likely to traverse edges of high strength and thus nodes
connected to node $s$ via paths of strong edges will likely be visited by the
random walk and will thus rank higher.

\xhdr{The optimization problem}
The training data contains information that source node $\s$ will create edges
to nodes $\d \in \D$ and not to nodes $\l \in \L$. So, we aim to set the
parameters $\prm$ of function $f_{\prm}(\atrij{uv})$ so that it will assign
edge weights $\wij{uv}$ in such a way that the random walk will be more likely
to visit nodes in $\D$ than $L$, \emph{i.e.}, $\p_{\l} < \p_{\d}$, for each $\d
\in \D$ and $\l \in \L$.

Thus, we define the optimization problem to find the optimal set of parameters
$\prm$ of edge strength function $f_{\prm}(\atrij{uv})$ as follows:
\begin{equation}
\begin{aligned}
    & \min_{\prm} F(\prm) = ||\prm||^2\\
    & \textrm{such that}\\
    & \forall \ \d{\in}\D, \l{\in}\L:\ \p_{\l} < \p_{\d}
\end{aligned}
\label{eq:hard}
\end{equation}
where $\p$ is the vector of PageRank scores. Note that PageRank scores $\p_i$
depend on edge strengths $\wij{uv}$ and thus actually depend on
$f_{\prm}(\atrij{uv})$ that is parameterized by $\prm$. The idea here is that
we want to find the parameter vector $\prm$ such that the PageRank scores of
nodes in $\D$ will be greater than the scores of nodes in $\L$. We prefer the
shortest $\prm$ parameter vector simply for regularization.

However, Eq.~\ref{eq:hard} is a ``hard'' version of the optimization problem as
it  allows no constraints to be violated. In practice it is unlikely that a
solution satisfying all the constraints exists. Thus similarly to formulations
of Support Vector Machines we make the constraints ``soft'' by introducing a
loss function $h$ that penalizes violated constraints. The optimization problem
now becomes:
\begin{equation}
\begin{aligned}
    & \min_{\prm} F(\prm) = ||\prm||^2 + \lambda \sum_{d \in D, l\in L} h(\p_{\l} - p_{\d})
\end{aligned}
\label{eq:opt}
\end{equation}
where $\lambda$ is the regularization parameter that trades-off between the
complexity (\emph{i.e.}, norm of $\prm$) for the fit of the model (\emph{i.e.},
how much the constraints can be violated). Moreover, $h(\cdot)$ is a loss
function that assigns a non-negative penalty according to the difference of the
scores $\p_{\l} - \p_{\d}$. If $\p_{\l} - \p_{\d} < 0$ then $h(\cdot)=0$ as
$\p_{\l} < \p_{\d}$ and the constraint is not violated, while for $\p_{\l} -
\p_{\d} > 0$, also $h(\cdot)>0$.

\xhdr{Solving the optimization problem}
First we need to establish the connection between the parameters $\prm$ of the
edge strength function $f_{\prm}(\atrij{uv})$ and the random walk scores $p$.
Then we show how to obtain the derivative of the loss function and the random
walk scores $p$ with respect to $\prm$ and then perform gradient based
optimization method to minimize the loss and find the optimal parameters
$\prm$.

Function $f_{\prm}(\atrij{uv})$ combines the attributes $\atrij{uv}$ and the
parameter vector $\prm$ to output a non-negative weight $\wij{uv}$ for each
edge.
%
%
%
We then build the random walk stochastic transition matrix $\Q'$:
\begin{equation}
\Q'_{uv}=
\begin{cases}
    \frac{\wij{uv}}{\sum_w \wij{uw}} & \text{if $(u,v) \in E$,} \\
    0 & \text{otherwise} \\
\end{cases}
\label{eq:qmatrix}
\end{equation}
To obtain the final random walk transition probability matrix $Q$, we also
incorporate the restart probability $\alpha$, \emph{i.e.}, with probability
$\alpha$ the random walk jumps back to seed node $s$ and thus ``restarts'':
$$
  Q_{uv} = (1-\alpha) Q'_{uv} + \alpha \mathbf{1}(v=s).
$$
Note that each row of $\Q$ sums to $1$ and thus each entry $\Q_{uv}$ defines
the conditional probability that a walk will traverse edge $(u,v)$ given that
it is currently at node $u$.

The vector $\p$ is the stationary distribution of the Random walk with restarts
(also known as Personalized PageRank), and is the solution to the following
eigenvector equation:
\begin{equation}
  \p^{T} = \p^{T}\Q \label{eq:pagerank}
\end{equation}

Equation~\ref{eq:pagerank} establishes the connection between the node PageRank
scores $\p_u \in \p$, and parameters $\prm$ of function $f_w(\atrij{uv})$ via
the random walk transition matrix $Q$. Our goal now is to minimize
Eq.~\ref{eq:opt} with respect to the parameter vector $\prm$.
%
We approach this by first deriving the gradient of $F(\prm)$ with respect to
$\prm$, and then use a gradient based optimization method to find $\prm$ that
minimize $F(\prm)$. Note that is non-trivial due to the recursive relation in
Eq.~\ref{eq:pagerank}.

First, we introduce a new variable $\delta_{\l\d} = \p_{\l} - \p_{\d}$ and then
we can write the derivative:
\begin{equation}
\begin{aligned}
    \deriv{\prm}{F(\prm)} & = & 2\prm + \sum_{l,d} \deriv{\prm}{h(\p_{\l} - \p_{\d})}  \\
                    & = & 2\prm + \sum_{l,d} \deriv{\delta_{\l\d}}{h(\delta_{\l\d})} (\deriv{\prm}{\p_{\l}} - \deriv{\prm}{\p_{\d}})
\end{aligned}
    \label{eq:optprob}
\end{equation}
For commonly used loss functions $h(\cdot)$ (like, hinge-loss or squared loss),
it is simple to compute the derivative
$\deriv{\delta_{\l\d}}{h(\delta_{\l\d})}$. However, it is not clear how to
compute $\deriv{\prm}{\p_{u}}$, the derivative of the score $\p_u$ with respect
to the vector $\prm$. Next we show how to do this.

Note that $\p$ is the principal eigenvector of matrix $\Q$.
Eq.~\ref{eq:pagerank} can be rewritten as $\p_u = \sum_j \p_j \Q_{ju}$ and
taking the derivative now gives:
\begin{equation}
    \deriv{\prm}{\p_u} = \sum_j \Q_{ju}\deriv{\prm}{\p_j} + \p_j\deriv{\prm}{\Q_{ju}}
    \label{eq:deriv}
\end{equation}
Notice that $\p_u$ and $\deriv{\prm}{\p_{u}}$ are recursively entangled in the
equation. However, we can still compute the gradient iteratively
~\cite{andrew79iterative,agarwal06rank}. By recursively applying the chain rule
to Eq.~\ref{eq:deriv} we can use a power-method like algorithm to compute the
derivative. We repeatedly compute the derivative $\deriv{\prm}{\p_u}$ based on
the estimate obtained in the previous iteration. Thus, we first compute $p$ and
then update the estimate of the gradient $\deriv{\prm}{\p_u}$. We stop the
algorithm when both $\p$ and $\deriv{\prm}{\p}$ do not change (\emph{i.e.},
$\varepsilon=10^{-12}$ in our experiments) between iterations. We arrive at
Algorithm~\ref{alg:deriv} that iteratively computes the eigenvector $\p$ as
well as the partial derivatives of $\p$. Convergence of
Algorithm~\ref{alg:deriv} is similar to those of
power-iteration~\cite{andrew78convergence}.

\begin{algorithm}[t]
    \dontprintsemicolon
    Initialize PageRank scores $\p$ and partial derivatives
    $\deriv{\prm_k}{\p_{u}}$:\\
    \Indp
    \lForEach{$u \in V$}{$\p_u^{(0)}=\frac{1}{|V|}$} \;
    \lForEach{$u \in V, k = 1, \dots, |\prm|$}{$\deriv{\prm_k}{\p_{u}}^{(0)}=0$} \;
    \Indm
    $t = 1$ \;
    \While{not converged}{
        \ForEach{$u \in V$}{
            $\p_u^{(t)} = \sum_{j} \p_j^{(t-1)} \Q_{ju}$\;
        }
        $t = t + 1$\;
    }
    $t = 1$ \;
    \ForEach{$k=1,\dots,|\prm|$}{
      \While{not converged}{
        \ForEach{$u \in V$}{
          $\deriv{\prm_k}{\p_u}^{(t)} = \sum_j \Q_{ju}\deriv{\prm_k}{\p_j}^{(t-1)} +
            \p_j^{(t-1)} \deriv{\prm_k}{\Q_{ju}}$\;
        }
        $t = t + 1$\;
      }
    }
    \Return $\deriv{\prm}{\p_u}^{(t-1)}$
 \caption{Iterative power-iterator like computation of PageRank vector $\p$ and its derivative $\deriv{\prm}{\p_u}$.}
 \label{alg:deriv}
\end{algorithm}

To solve Eq.~\ref{eq:pagerank} we further need to compute
$\deriv{\prm}{\Q_{ju}}$ which is the partial derivative of entry $\Q_{ju}$
(Eq.~\ref{eq:qmatrix}). This calculation is straightforward. When $(j,u) \in E$ we
find $\deriv{\prm}{\Q_{ju}} =$
$$
    (1-\alpha) \frac{\deriv{\prm}{f_{\prm}(\atrij{ju})}\big(\sum_k f_{\prm}(\atrij{jk})\big)
        - f_{\prm}(\atrij{ju})\big(\sum_k \deriv{\prm}{f_{\prm}(\atrij{jk})} \big)}
    {\big(\sum_k f_{\prm}(\atrij{jk})\big)^2}
$$
and otherwise $\deriv{\prm}{\Q_{ju}} =0$. The edge strength function
$f_{\prm}(\atrij{uv})$ must be differentiable and so
$\deriv{\prm}{f_{\prm}}(\atrij{jk})$ can be easily computed.

This completes the derivation and shows how to evaluate the derivative of
$F(\prm)$ (Eq.~\ref{eq:optprob}). Now we apply a gradient descent based method,
like a quasi-Newton method, and directly minimize $F(\prm)$.

\xhdr{Final remarks} First we note that our problem is not convex in general,
and thus gradient descent methods will not necessarily find the global minimum.
In practice we resolve this by using several different starting points to find a
good solution.

Second, since we are only interested in the values of $\p$ for nodes in $C$, it
makes sense to evaluate the loss function at a slightly different point:
$h(\p'_l-\p'_d)$ where $\p'$ is a normalized version of $\p$ such that $\p'_u =
\frac{\p_u}{\sum_{v \in C} \p_v}$. This adds one more chain rule application to
the derivative calculation, but does not change the algorithm. The effect of
this is mostly to allow larger values of $\alpha$ to be used without having to
change $h(\cdot)$
(We omit the tick marks in our notation for the rest of this paper, using $\p$
to refer to the normalized score).

So far we only considered training and estimating the parameter vector $\prm$
for predicting the edges of a particular node $s$. However, our aim to estimate
$\prm$ that make good predictions across many different nodes $s \in S$. We
easily extend the algorithm to multiple source nodes $s \in S$, that may even
reside in different graphs. We do this by taking the sum of losses over all
source nodes $s$ and the corresponding pairs of positive $D_s$ and negative
$L_s$ training examples. We slightly modify the Eq.~\ref{eq:opt} to obtain:

$$ \min_{\prm} F(\prm) = ||\prm||^2 + \lambda \sum_{s\in S} \sum_{d \in D_s,
l\in L_s} h(\p_{\l} - p_{\d}) $$

\noindent The gradients of each instance $s \in S$ remain independent, and can
thus be computed independently for all instances of $s$ (Alg.~\ref{alg:deriv}).
By optimizing parameters $\prm$ over many individuals $s$, the algorithm is
less likely to overfit, which improves the generalization.

As a final implementation note, we point out that gradient descent often makes
many small steps which have small impact on the eigenvector and its derivative.
A 20\% speedup can be achieved by using the solutions from the previous
position (in the gradient descent) as initialization for the eigenvector and
derivative calculations in Alg.~\ref{alg:deriv}.  Our implementation of
Supervised Random Walks uses the L-BFGS algorithm~\cite{lbfgs}. Given a
function and its partial derivatives, the solver iteratively improves the
estimate of $\prm$, converging to a local optima. The exact runtime of the
method depends on how many iterations are required for convergence of both the
PageRank and derivative computations, as well as of the overall process
(quasi-Newton iterations).

\vspace{1cm}
\section{Experiments on synthetic data}

Before experimenting with real data, we examine the soundness and robustness of
the proposed algorithm using synthetic data. Our goal here is to generate
synthetic graphs, edge features and training data (triples $(s, D, L)$) and
then try to recover the original model.

\xhdr{Synthetic data} We generate scale-free graphs $G$ on 10,000 nodes by
using the Copying model~\cite{kumar00stochastic}: Graph starts with three nodes
connected in a triad. Remaining nodes arrive one by one, each creating exactly
three edges. When a node $u$ arrives, it adds three edges $(u,v_i)$. Existing
node $v_i$ is selected uniformly at random with probability $0.8$, and
otherwise $v_i$ is selected with probability proportional to its current
degree. For each edge $(u,v)$ we create two independent Gaussian features with
mean $0$ and variance $1$. We set the edge strength $\wij{uv} = exp(\atrij{uv1}
- \atrij{uv2})$, \emph{i.e.}, $\prm^*=[1,-1]$.

For each $G$, we randomly select one of the oldest 3 nodes of $G$ as the start
node, $s$. To generate a set of destination $D$ and no-link nodes $L$ for a
given $s$ we use the following approach.

On the graph with edge strengths $\wij{uv}$ we run the random walk
($\alpha=0.2$) starting from $s$ and obtain node PageRank scores $\p^*$. We use
these scores to generate the destinations $D$ in one of two ways. First is
deterministic and selects the top $K$ nodes according to $\p^*$ to which $s$ is
not already connected. Second is probabilistic and selects $K$ nodes, selecting
each node $u$ with probability $\p^*_u$.

Now given the graph $G$, attributes $\atrij{uv}$ and targets $D$ our goal is to
recover the true edge strength parameter vector $\prm^*=[1,-1]$. To make the
task more interesting we also add random noise to all of the attributes, so
that $\atrij{uvi}' = \atrij{uvi} + \mathcal{N}(0,\sigma^2)$, where
$\mathcal{N}(0,\sigma^2)$ is a Gaussian random variable with mean 0 and
variance $\sigma^2$.

\begin{figure}[t]
    \centering
    \includegraphics[width=5.3cm,angle=270]{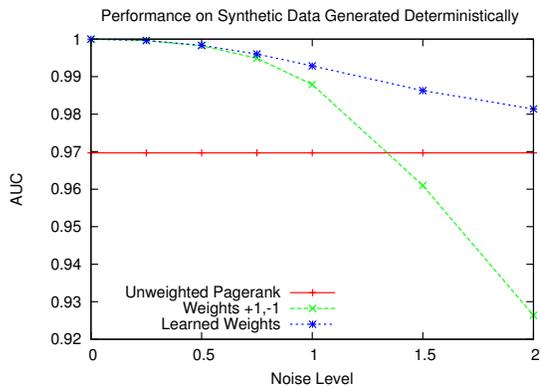}
    \vspace{-3mm}
    \caption{Experiments on synthetic data. Deterministic $D$.}
    \vspace{-3mm}
    \label{fig:synthetic-det}
\end{figure}

\begin{figure}[t]
    \centering
    \includegraphics[width=5.3cm,angle=270]{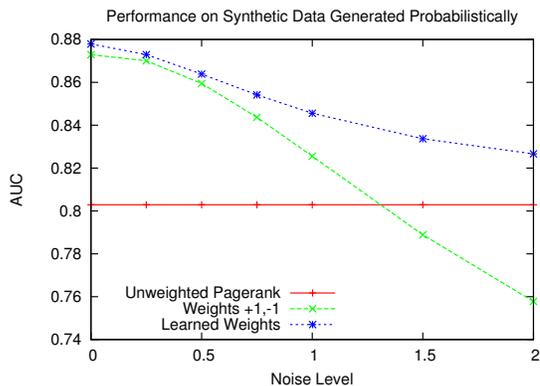}
    \vspace{-3mm}
    \caption{Experiments on synthetic data. Probabilistic $D$.}
    \vspace{-3mm}
    \label{fig:synthetic-prob}
\end{figure}

\xhdr{Results} After applying our algorithm, we are interested in two things.
First, how well does the model perform in terms of the classification accuracy
and second, whether it recovers the edge strength function parameters
$\prm^*=[1,-1]$. In the deterministic case of creating $D$ and with $0$ noise
added, we hope that the algorithm is able achieve near perfect classification.
As the noise increases, we expect the performance to drop, but even then, we
hope that the recovered values of $\hat{\prm}$ will be close to true $\prm^*$.

In running the experiment we generated 100 synthetic graphs.  We used 50 of
them for training the weights $\prm$, and report results on the other 50.  We
compute Area under the ROC curve (AUC) of each of 50 test graphs, and report
the mean (AUC of 1.0 means perfect classification, while random guessing scores
0.5).

Figures~\ref{fig:synthetic-det} and~\ref{fig:synthetic-prob} show the results.
We plot the performance of the model that ignores edge weights (red), the model
with true weights $\prm^*$ (green) and a model with learned weights
$\hat{\prm}$ (blue).

For the deterministically generated $D$ (Fig.~\ref{fig:synthetic-det}), the
performance is perfect in the absence of any noise. This is good news as it
demonstrates that our training procedure is able to recover the correct
parameters. As the noise increases, the performance slowly drops.  When the
noise reaches $\sigma^2 \approx 1.5$, using the true parameters $w^*$ (green)
actually becomes worse than simply ignoring them (red). Moreover, our algorithm
learns the true parameters $[+1,-1]$ almost perfectly in the noise-free case,
and decreases their magnitude as the noise level increases. This matches the
intuition that, as more and more noise is added, the signal in the edge
attributes becomes weaker and weaker relatively to the signal in the graph
structure. Thus, with more noise, the parameter values $\prm$ decrease as they
are given less and less credence.

In the probabilistic case (Fig.~\ref{fig:synthetic-prob}), we see that our
algorithm does better (statistically significant at $p=0.01$) than the model
with true parameters $\prm^*$, regardless of the presence or absence of noise.
Even though the data was generated using parameters $w^*=[+1,-1]$, these values
are not optimal and our model gets better AUC by finding different (smaller)
values. Again, as we add noise, the overall performance slowly drops, but still
does much better than the baseline method of ignoring edge strengths (red), and
continues to do better than the model that uses true parameter values $w^*$
(green).

We also note that regardless of where we initialize the parameter vector $\prm$
before starting gradient descent, it always converges to the same solution.
Having thus validated our algorithm on synthetic data, we now move on to
predicting links in real social networks.

\section{Experimental setup}
\label{sec:experiments}

\begin{figure}[t]
    \centering
    \includegraphics[width=6cm,angle=270]{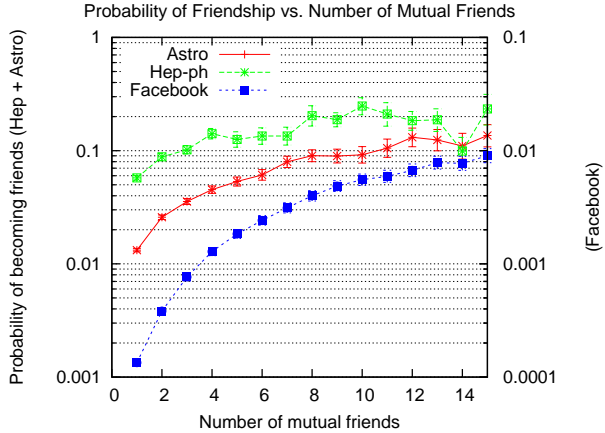}
    \vspace{-3mm}
    \caption{Probability of a new link as a function of the number of mutual friends.}
    \vspace{-3mm}
    \label{fig:pvmf}
\end{figure}

\begin{figure}[t]
    \centering
    \includegraphics[width=6cm]{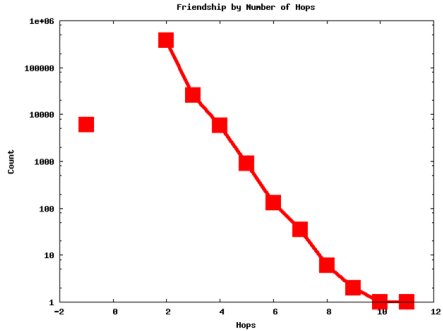}
    \vspace{-3mm}
    \caption{Facebook Iceland:
    Hop distance between a pair of nodes just before they become friends.
    Distance x=-1 denotes nodes that were in separate components, while x=2 (friends
    of friends) is order of magnitude higher than next highest point.}
    \vspace{-3mm}
    \label{fig:fbdist}
\end{figure}

\begin{table}[t]
  \centering
  \begin{tabular}{l||r|r|r|r|r|r}
    & $N$ & $E$ & S & $\bar{D}$ & $\bar{C}$ & $\bar{D}/\bar{C}$ \\ \hline \hline
    Astro-Ph & 19,144 & 198,110 & 1,123 & 18.0 & 775.6 &  0.023 \\ \hline
    Cond-Mat  & 23,608 & 94,492 & 140 & 9.1 & 335.5 & 0.027 \\ \hline
    Hep-Ph  &  12,527 & 118,515 & 340 & 29.2 & 345.3 & 0.084 \\ \hline
    Hep-Th  & 10,700 & 25,997 & 55 & 6.3 & 110.5 &  0.057 \\ \hline \hline
    Facebook  & 174,000 & 29M & 200 & 43.6 & 1987 & 0.022
  \end{tabular}
\vspace{-3mm}
  \caption{Dataset statistics. $N, E$: number of nodes and edges in
  the full network, $S$: number of sources, $\bar{C}$: avg. number of candidates
  per source, $\bar{D}$: avg. number of destination nodes.}
  \label{tab:datasets}
\vspace{-3mm}
\end{table}

For experiments on real data we consider four real physics co-authorship
networks and a complete Facebook network of Iceland.

Generally we focus on predicting links to nodes that are 2-hops from the seed
node $s$. We do this for two reasons. First, in online social networks more
than half of all edges at the time of creation close a triangle, \emph{i.e.}, a
person connects to a friend of a friend~\cite{jure08microevol}. For instance,
Figure~\ref{fig:fbdist} shows that 92\% of all edges created on Facebook
Iceland close a path of length two, \emph{i.e.}, a triangle. Second, this also
makes the Supervised Random Walks run faster as graphs get smaller. Given that
some Facebook users have degrees in the thousands, it is not practical to
incorporate them (a user may have as many as a hundred million nodes at 3
hops).

\xhdr{Co-authorship networks} First we consider the co-authorship networks from
arXiv e-print archive~\cite{gehrke03kddcup} where we have a time-stamped list
of all papers with author names and titles submitted to arXiv during 1992 and
2002. We consider co-authorship networks from four different areas of physics:
Astro-physics (Astro-Ph), Condensed Matter (Cond-Mat), High energy physics
theory (Hep-th) and High energy physics phenomenology (Hep-ph). For each of the
networks we proceed as follows. For every node $u$ we compute the total number
of co-authors at the end of the dataset (\emph{i.e.}, network degree) $k_u$ and
let $t_u$ be the time when $u$ created it's $k_u/2$-th edge. Then we define
$m_u$ to be the number of co-authorship links that $u$ created after time $t_u$
and that at the time of creation spanned 2-hops (\emph{i.e.}, closed a
triangle). We attempt to make predictions only for ``active'' authors, where we
define a node $u$ to be active if $k_u \ge K$ and $m_u \ge \Delta$. In this
work, we set $K=10$ and $\Delta=5$.
For every source node $s$ that is above this threshold, we extract the network
at time $t_s$ and try to predict the $d_s$ new edges that $s$ creates in the
time after $t_s$. Table~\ref{tab:datasets} gives dataset statistics.

For every edge $(i,j)$ of the network around the source node $u$ at time $t_u$
we generate the following six features:
\begin{itemize}
  \denselist
  \item Number of papers $i$ written before $t_u$
  \item Number of papers $j$ written before $t_u$
  \item Number of papers $i$ and $j$ co-authored
  \item Cosine similarity between the titles of papers written by $i$ and
      titles of $j$'s papers
  \item Time since $i$ and $j$ last co-authored a paper.
  \item The number of common friends between $j$ and $s$.
\end{itemize}

\xhdr{The Facebook network} Our second set of data comes from the Facebook
online social network.  We first selected Iceland since it has high Facebook
penetration, but relatively few edges pointing to users in other countries. We
generated our data based on the state of the Facebook graph on November 1,
2009. The destination nodes $\D$ from a node $s$ are those that $s$ became
friends with between November 1 2009 and January 13 2010.
The Iceland graph contains more than 174 thousand people, or 55\% of the
country's population.  The average user had 168 friends, and during the period
Nov 1 -- Jan 23, an average person added 26 new friends.

From these users, we randomly selected 200 as the nodes $s$.  Again, we only
selected ``active'' nodes, this time with the criteria $|\D| > 20$.
As Figure~\ref{fig:pvmf} shows, individuals without many mutual friends are
exceedingly unlikely to become friends. As the Facebook graph contains users
whose 2-hop neighborhood have several million nodes we can prune such graphs
and speed-up the computations without loosing much on prediction performance.
Since we know that individuals with only a few mutual friends are unlikely to
form friendships, and our goal is to predict the most likely friendships, we
remove all individuals with less than 4 mutual friends with practically no loss
in performance.
As demonstrated in Figure~\ref{fig:pvmf}, if a user creates an edge, then the
probability that she links to a node with whom she has less than 4 friends is
about $0.1\%$.).

We annotated each edge of the Facebook network with seven features. For each
edge $(i,j)$, we created:
\begin{itemize}
  \denselist
  \item Edge age: $(T-t)^{-\beta}$, where $T$ is the time cutoff Nov. 1,
      and $t$ is the edge creation time.  We create three features like
      this with $\beta = \{0.1,0.3,0.5\}$.
  \item Edge initiator: Individual making the friend request is encoded as
      $+1$ or $-1$.
  \item Communication and observation features.  They represent the
      probability of communication and profile observation in a one week
      period. 
  \item The number of common friends between $j$ and $s$.
  
\end{itemize}
All features in all datasets are re-scaled to have mean 0 and standard
deviation 1. We also add a constant feature with value 1.

\xhdr{Evaluation methodology} For each dataset, we assign half of the nodes $s$
into training and half into test set. We use the training set to train the
algorithm (\emph{i.e.}, estimate $\prm$). We evaluate the method on the test
set, considering two performance metrics: the Area under the ROC curve (AUC)
and the Precision at Top 20 (Prec@20), \emph{i.e.}, how many of top 20 nodes
suggested by our algorithm actually receive links from $s$. This measure is
particularly appropriate in the context of link-recommendation where we present
a user with a set of friendship suggestions and aim that most of them are
correct.

\section{Experiments on real data}

Next we describe the results of on five real datasets: four co-authorship
networks and the Facebook network of Iceland.

\subsection{General considerations}

First we evaluate several aspects of our algorithm: (A) the choice of the loss
function, (B) the choice of the edge strength function $f_w(\cdot)$, (C) the
choice of random walk restart (jump) parameter $\alpha$, and (D) choice of
regularization parameter $\lambda$. We also consider the extension where we
learn a separate edge weight vector depending on the type of the edge,
\emph{i.e.}, whether an edge touches $s$ or any of the candidate nodes $c \in
C$.

\xhdr{(A) Choice of the loss function} As is the case with most machine
learning algorithms, the choice of loss function plays an important role.
Ideally we would like to optimize the loss function $h(\cdot)$ which directly
corresponds to our evaluation metric (\emph{i.e.}, AUC or Precision at top
$k$). However, as such loss functions are not continuous and not differentiable
and so it is not clear how to optimize over them. Instead, we experiment with
three common loss functions:

\begin{itemize}
  \denselist
  \item Squared loss with margin $b$: $$ h(x) = \max\{x + b,0\}^2
  $$
  \item Huber loss with margin $b$ and window $z > b$:
    \begin{equation}
    h(x)=
    \begin{cases}
        0 & \text{if $x \le -b$,} \\
        (x+b)^2/(2z) & \text{if $-b < x \le z-b$,} \\
        (x+b) - z/2 & \text{if $x > z-b$} \\
    \end{cases}
    \label{eq:huber}
    \end{equation}
  \item Wilcoxon-Mann-Whitney (WMW) loss with width $b$ (Proposed to be used when
  one aims to maximize AUC~\cite{yan03logistic}):
    $$h(x) = \frac{1}{1 + exp(-x / b)}$$
\end{itemize}

Each of these loss functions is differentiable and needs to be evaluated for
all pairs of nodes $d \in D$ and $l \in L$ (see Eq.~\ref{eq:opt}). Performing
this naively takes approximately $O(c^2)$ where $c = |D \cup L|$. However, we
next show that the first two loss functions have the advantage that they can be
computed in $O(c \log c)$. For example, we rewrite the squared loss as:
\begin{equation}
\begin{aligned}
  \sum_{d, l} h(\p_\l - \p_\d)  & = & \sum_{\l,\d : \p_\l + b > \p_\d} (\p_\l-\p_\d+b)^2 \\
  & = & \sum_l \sum_{\d : \p_\l + \b > \p_\d} (\p_\l+b)^2 - 2 (\p_\l + b) \p_\d + \p_\d^2 \\
  & = & \sum_l |\{\d : \p_\l + \b> \p_\d\}| (\p_\l+b)^2 \\
  &  & - 2 (\p_\l + b)\sum_{d : \p_\l + \b > \p_\d} \p_\d + \sum_{d : \p_\l + \b> \p_\d} \p_\d^2
    \nonumber
\end{aligned}
\end{equation}
Once we have the lists $\{\p_\l\}$ and $\{\p_\d\}$ sorted, we can iterate over
the list $\{\p_\l\}$ in reverse order.  As we do this, we can incrementally
update the two terms which sum over $\d$ above. The Huber loss can as well be
quickly evaluated using a similar calculation.

Computation of the WMW loss is more expensive, as there is no way to go around
the summation over all pairs. Evaluating WMW loss thus takes time $O(|\D| \cdot
|\L|)$.  In our case, $|\D|$ is typically relatively small, and so the
computation is not a significant part of total runtime. However, the primary
advantage of it is that it performs slightly better. Indeed, in the limit as
$b$ goes to 0, it reflects AUC, as it measures the number of inversions in the
ordering~\cite{yan03logistic}.

In our experiments we notice that while the gradient descent achieves
significant reduction in the value of the loss for all three loss functions,
this only translates to improved AUC and Prec@20 for the WMW loss.  In fact,
the model trained with the squared or the Huber loss does not perform much
better than the baseline we obtain through unweighted PageRank. Consequently,
we use the WMW loss function for the remainder of this work.

\xhdr{(B) Choice of edge strength function $\bf f_w(\atrij{uv})$} The edge
strength function $f_w(\atrij{uv})$ must be non-negative and differentiable.
While more complex functions are certainly possible, we experiment with two
functions.  In both cases, we start by taking the inner product of the weight
vector $\prm$ and the feature vector $\atrij{uv}$ of an edge $(u,v)$. This
yields a single scalar value, which may be negative. To transform this into the
desired domain, we apply either an exponential or logistic function:

\begin{itemize}
\denselist
\item Exponential edge strength: $\wij{uv} = \exp(\atrij{uv} \cdot \prm)$
\item Logistic edge strength: $\wij{uv} = (1 + \exp(- \atrij{uv} \cdot
    \prm))^{-1}$
\end{itemize}

Our experiments show that the choice of the edge strength function does not
seem to make a significant impact on performance.  There is slight evidence
from our experiments that the logistic function performs better 
One problem that can occur with the exponential version is underflow and
overflow of double precision floating point numbers. As the performance seems
quite comparable, we recommend the use of the logistic to avoid this potential
pitfall.

\begin{figure}
    \centering
    \includegraphics[width=5.2cm,angle=270]{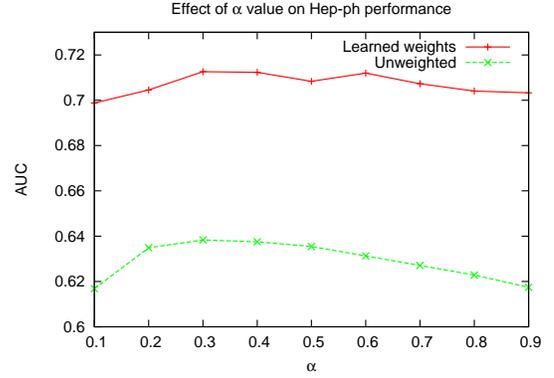}
    \vspace{-3mm}
    \caption{Impact of random walk restart parameter $\bf \alpha$.}
    \label{fig:alpha}
    \vspace{-3mm}
\end{figure}

\xhdr{(C) Choice of $\bf \alpha$} To get a handle on the impact of random walk
restart parameter $\alpha$, it is useful to think of the extreme cases, for
unweighted graphs. When $\alpha=0$, the PageRank of a node in an undirected
graph is simply its degree. On the other hand, when $\alpha$ approaches 1, the
score will be exactly proportional to the ``Random-Random''
model~\cite{jure08microevol} which simply makes two random hops from $s$, as
random walks of length greater than 2 become increasingly unlikely, and hence
the normalized eigenvector scores become the same as the Random-Random
scores~\cite{jure08microevol}.
When we add the notion of edge strengths, these properties remain. Intuitively,
$\alpha$ controls for how ``far'' the walk wanders from seed node $s$ before it
restarts and jumps back to $s$. High values of $\alpha$ give very short and
local random walks, while low values allow the walk to go farther away.

When evaluating on real data we observe that $\alpha$ plays an important role
in the simple unweighted case when we ignore the edge strengths, but as we give
the algorithm more power to assign different strengths to edges, the role of
$\alpha$ diminishes, and we see no significant difference in performance for a
broad range of choices $\alpha$. Figure~\ref{fig:alpha} illustrates this; in
the unweighted case (\emph{i.e.}, ignoring edge strengths) $\alpha=0.3$
performs best, while in the weighted case a broad range from $0.3$ to $0.7$
seem to do about equally well.

\xhdr{(D) Regularization parameter $\bf \lambda$} Empirically we find that
overfitting is not an issue in our model as the number of parameters $\prm$ is
relatively small. Setting $\lambda=1$ gives best performance.

\xhdr{Extension: Edge types} The Supervised Random Walks framework we have
presented so far captures the idea that some edges are stronger than others.
However, it doesn't allow for different types of edges.  For instance, it might
be that an edge $(u,v)$ between $s$'s friends $u$ and $v$ should be treated
differently than the edge $(s,u)$ between $s$ and $u$.  Our model can easily
capture this idea by declaring different edges to be of different types, and
learning a different set of feature weights $\prm$ for each edge type.  We can
take the same approach to learning each of these weights, computing partial
derivatives with respect to each one weight. The price for this is potential
overfitting and slower runtime.

In our experiments, we find that dividing the edges up into multiple types
provides significant benefit.  
Given a seed node $s$ we label the edges according to the hop-distance from $s$
of their endpoints, \emph{e.g.}, edges $(s,u)$ are of type (0,1), edges $(u,v)$
are either of type (1,1) (if both $u$ and $v$ link to $s$) or (1,2) (if $v$
does not link to $s$). Since the nodes are at distance 0, 1, or 2 from $s$,
there are 6 possible edge types: (0,1), (1,0), (1,1), (1,2), (2,1) and (2,2).
While learning six sets of more parameters $\prm$ increases the runtime, using
multiple edge types gives a significant increase in performance.

\begin{figure}[t]
    \centering
    \includegraphics[width=5.2cm]{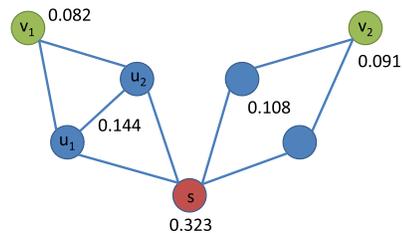}
    \vspace{-3mm}
    \caption{Stationary random walk distribution with $\alpha = 0.15$.}
    \label{fig:pagerank}
    \vspace{-3mm}
\end{figure}

\xhdr{Extension: Social capital} Before moving on to the experimental results,
we also briefly examine somewhat counterintuitive behavior of the Random Walk
with Restarts. Consider a graph in Figure~\ref{fig:pagerank} with the seed node
$s$.  There are two nodes which $s$ could form a new connection to $v_1$ and
$v_2$.  These two are symmetric except for the fact that the two paths
connecting $s$ to $v_1$ are connected themselves. Now we ask, is $s$ more
likely to link to $v_1$ or to $v_2$?

Building on the theory of embeddedness and social
capital~\cite{coleman88capital} one would postulate that $s$ is more likely to
link to $v_1$ than to $v_2$. However, the result of an edge $(u_1,u_2)$ is that
when $\alpha > 0$, $v_2$ ends up with a higher PageRank score than $v_1$.  This
is somewhat counterintuitive, as $v_1$ somehow seems ``more connected'' to $s$
than $v_2$. Can we remedy this in a natural way?

One solution could be that carefully setting $\alpha$ resolves the issue.
However, there is no value of $\alpha>0$ which will make the score of $v_1$
higher than $v_2$ and changing to other simple teleporting schemes (such as a
random jump to a random node) does not help either.  However, a simple
correction that works is to add the number of friends a node $w$ has in common
with $s$, and use this as an additional feature $\gamma$ on each edge $(u,w)$.
If we apply this to the graph shown in Figure~\ref{fig:pagerank}, and set the
weight along each edge to $1+\gamma$, then the PageRank score  $\p_{v_1}$ of
node $v_1$ is 1.9 greater than of $v_2$ (as opposed to 0.1 smaller as in
Fig~\ref{fig:pagerank}).

In practice, we find that introducing this additional feature $\gamma$ helps on
the Facebook graph. In Facebook, connection $(u_1,u_2)$ increases the
probability of a link forming to $v_1$ by about 50\%. In the co-authorship
networks, the presence of $(u_1,u_2)$ actually decreases the link formation
probability by 37\%. 
Such behavior of co-authorship networks can be explained by the argument that
long range weak ties help in access to new information~\cite{granovetter73ties}
(\emph{i.e.}, $s$ is more likely to link to $v_2$ than $v_1$ of
Fig~\ref{fig:pagerank}). Having two independent paths is a stronger connection
in the co-authorship graph, as this indicates that $s$ has written papers with
two people, on two different occasions, and both of these people have written
with the target $v$, also on two different occasions. Thus, there must be at
least four papers between these four people when the edge $(u_1,u_2)$ is
absent, and there may be as few as two when it is present. Note this is exactly
the opposite to the social capital argument~\cite{coleman88capital}, which
postulates that individuals who are well embedded in a network or a community
have higher trust and get more support and information. This is interesting as
it shows that Facebook is about social contacts, norms, trying to fit in and be
well embedded in a circle of friends, while co-authorship networks are about
access to information and establishing long-range weak ties.

\subsection{Experiments on real data}

Next we evaluate the predictive performance of Supervised Random Walks (SRW) on
real datasets. We examine the performance of the parameter estimation and then
compare Supervised Random Walks to other link-prediction methods.

\begin{figure}[t]
    \centering
    \includegraphics[width=6cm,angle=270]{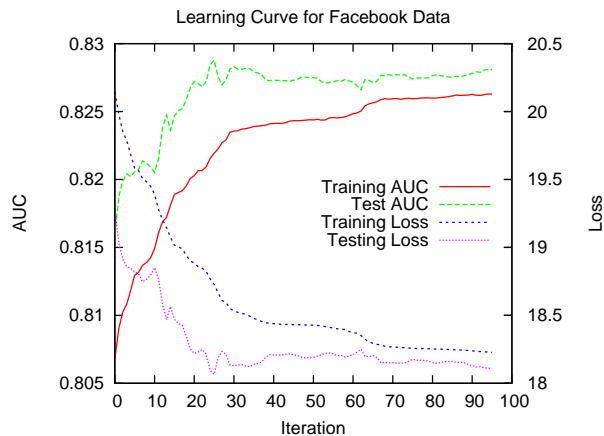}
    \vspace{-3mm}
    \caption{Performance of Supervised Random Walks as a function of the
    number of steps of parameter estimation procedure.}
    \label{fig:lc}
    \vspace{-3mm}
\end{figure}

\xhdr{Parameter estimation} Figure~\ref{fig:lc} shows the results of gradient
descent on the Facebook dataset.  At iteration 0, we start with unweighted
random walks, by setting $\prm=0$.  Using L-BFGS we perform gradient descent on
the WMW loss. Notice the strong correlation between AUC and WMW loss,
\emph{i.e.}, as the value of the loss decreases, AUC increases. We also note
that the method basically converges in only about 25 iterations.

\begin{table}[t]
  \centering
  \begin{tabular}{l||r|r}
    Learning Method & AUC & Prec@20 \\
    \hline \hline
    Random Walk with Restart & 0.63831 & 3.41 \\ 
    Adamic-Adar  & 0.60570 & 3.13 \\
    Common Friends & 0.59370 & 3.11 \\
    Degree & 0.56522 & 3.05 \\
    \hline
    DT: Node features  & 0.60961 & 3.54 \\
    DT: Network features & 0.59302 & 3.69 \\
    DT: Node+Network & 0.63711 & 3.95 \\
    DT: Path features  & 0.56213 & 1.72 \\
    DT: All features  & 0.61820 & 3.77 \\
    \hline
    LR: Node features  & 0.64754 & 3.19  \\
    LR: Network features & 0.58732 & 3.27 \\
    LR: Node+Network & 0.64644 & 3.81 \\
    LR: Path features  & 0.67237 & 2.78 \\
    LR: All features  & 0.67426 & 3.82 \\
    \hline
    SRW: one edge type  & 0.69996 & 4.24 \\
    SRW: multiple edge types   & {\bf 0.71238} & {\bf 4.25}
  \end{tabular}
  \vspace{-2mm}
  \caption{Hep-Ph co-authorship network.
  DT: decision tree, LR: logistic regression, and
  SRW: Supervised Random Walks.}
  \label{tab:realdata-hep}
  \vspace{-2mm}
\end{table}

\begin{table}[t]
  \centering
  \begin{tabular}{l||r|r}
    Learning Method & AUC & Prec@20 \\
    \hline \hline
    Random Walk with Restart & 0.81725 & 6.80 \\
    Adamic-Adar  & 0.81586 & 7.35 \\
    Common Friends & 0.80054 & 7.35 \\
    Degree & 0.58535 & 3.25 \\
    \hline
    DT: Node features  & 0.59248 & 2.38 \\
    DT: Network features & 0.76979 & 5.38 \\
    DT: Node+Network & 0.76217 & 5.86 \\
    DT: Path features  & 0.62836 & 2.46 \\
    DT: All features  & 0.72986 & 5.34 \\
    \hline
    LR: Node features  & 0.54134 & 1.38 \\
    LR: Network features & 0.80560 & 7.56 \\
    LR: Node+Network & 0.80280 & 7.56 \\
    LR: Path features  & 0.51418 & 0.74 \\
    LR: All features  & 0.81681 & 7.52 \\
    \hline
    SRW: one edge type  & {\bf 0.82502} & 6.87 \\
    SRW: multiple edge types   & {\bf 0.82799} & {\bf 7.57}
  \end{tabular}
  \vspace{-2mm}
  \caption{Results for the Facebook dataset.}
  \label{tab:realdata-fb}
  \vspace{-2mm}
\end{table}

\begin{table}[t]
  \centering
  \begin{tabular}{l||r|r||r|r}
    Dataset & \multicolumn{2}{c||}{AUC} & \multicolumn{2}{c}{Prec@20} \\
    & SRW & LR & SRW & LR \\
    \hline \hline
    Co-authorship Astro-Ph & {\bf 0.70548} & 0.67639 & 2.55 & 2.15 \\
    Co-authorship Cond-Mat & {\bf 0.74173} & 0.71672 & 2.54 & 2.61 \\
    Co-authorship Hep-Ph & {\bf 0.71238} & 0.67426 & 4.18 & 3.82 \\
    Co-authorship Hep-Th & {\bf 0.72505} & 0.69428 & 2.59 & 2.61 \\
    Facebook (Iceland) & {\bf 0.82799} & 0.81681 & 7.57 & 7.52
  \end{tabular}
  \vspace{-2mm}
  \caption{Results for all datasets.  We compare favorably to logistic
features as run on all features. Our Supervised Random Walks (SRW) perform
significantly better than the baseline in all cases on ROC area.  The variance
is too high on the Top20 metric, and the two methods are statistically tied on
this metric.}
  \label{tab:realdata-all}
  \vspace{-2mm}
\end{table}

\begin{figure}[t]
    \centering
    \includegraphics[width=5.2cm,angle=270]{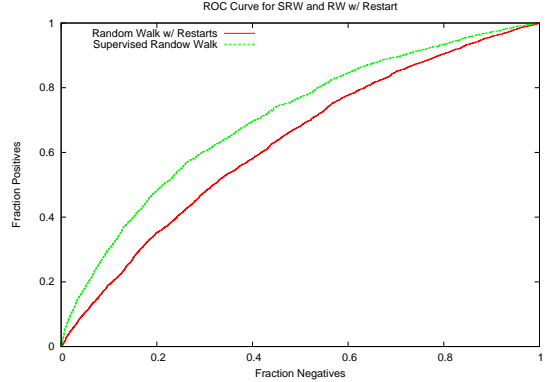}
    \vspace{-3mm}
    \caption{ROC curve of Astro-Ph test data.}
    \vspace{-3mm}
        \label{fig:as.roc}
\end{figure}

\xhdr{Comparison to other methods} Next we compare the predictive performance
of Supervised Random Walks (SRW) to a number of simple unsupervised baselines,
along with two supervised machine learning methods. All results are evaluated
by creating two independent datasets, one for training and one for testing.
Each performance value is the average over all of the graphs in the test set.

Figure~\ref{fig:as.roc} shows the ROC curve for Astro-Ph dataset, comparing our
method to an unweighted random walk.  Note that much of the improvement in the
curve comes in the area near the origin, corresponding to the nodes with the
highest predicted values. This is the area that we most care about,
\emph{i.e.}, since we can only display/recommend about 20 potential target
nodes to a Facebook user we want the top of the ranking to be particularly good
(and do not care about errors towards the bottom of the ranking).

We compare the Supervised Random Walks to unsupervised link-prediction methods:
plain Random Walk with Restarts, Adamic-Adar score~\cite{adamic03}, number of
common friends, and node degree. For supervised machine learning methods we
experiments with decision trees and logistic regression and group the features
used for training them into three groups:
\begin{itemize}
\denselist
  \item Network features:  unweighted random walk scores, Adamic-Adar
      score, number of common friends, and degrees of nodes $s$ and the
      potential target $c \in C$
  \item Node features: average of the edge features for those edges
      incident to the nodes $s$ and $c \in C$, as described in
      Section~\ref{sec:experiments}
  \item Path features: averaged edge features over all paths between seed
      $s$ and the potential destination $c$.
\end{itemize}


Tables~\ref{tab:realdata-hep} and~\ref{tab:realdata-fb} compare the results of
various methods on the Hep-Ph co-authorship and Facebook networks. In general,
we note very performance of  Supervised Random Walks (SRW): AUC is in the range
0.7--0.8 and precision at top 20 is between 4.2--7.6. We consider this
surprisingly good performance. For example, in case of Facebook this means that
out of 20 friendships we recommend nearly 40\% of them realize in near future.

Overall, Supervised Random Walks (SRW) give a significant improvement over the
unweighted Random Walk with Restarts (RWR).  SRW also gives gains over other
techniques such as logistic regression which combine features. For example, in
co-authorship network (Tab.~\ref{tab:realdata-hep}) we note that unsupervised
RWR outperforms decision trees and slightly trails logistic regression in terms
of AUC and Prec@20. Supervised Random Walks outperform all methods. In terms of
AUC we get 6\% and in terms of Prec@20 near 12\% relative improvement. In
Facebook (Tab.~\ref{tab:realdata-fb}), Random Walk with Restarts already gives
near-optimal AUC, while Supervised Random Walks still obtain 11\% relative
improvement in Prec@20.

It is important to note that, in addition to outperforming the other methods,
Supervised Random Walks do so without the tedious process of feature
extraction. There are many network features relating pairs of unconnected nodes
(Adamic-Adar was the best out of the dozens examined
in~\cite{libennowell03linkpred}, for example). Instead, we need only select the
set of node and edge attributes, and Supervised Random Walks take care of
determining how to combine them with the network structure to make predictions.

Last, Table~\ref{tab:realdata-all} compares the performance of top two methods:
Supervised Random Walks and logistic regression. We note that Supervised Random
Walks compare favorably to logistic regression. As logistic regression requires
state of the art network feature extraction and Supervised Random Walks
outperforms it out of the box and without any ad hoc feature engineering.

When we examine the weights assigned, we find that for Facebook the largest
weights are those which are related to time.  This makes sense as if a user has
just made a new friend $u$, she is likely to have also recently met some of
$u$'s friends.  In the co-authorship networks, we find that the number of
co-authored papers and the cosine similarity amongst titles were the features
with highest weights.

\xhdr{Runtime} While the exact runtime of Supervised Random Walks is highly
dependent on the graph structure and features used, we give some rough
guidelines. The results here are for single runs on a single 2.3Ghz processor
on the Facebook dataset.

When putting all edges in the same category, we have 8 weights to learn.  It
took 98 iterations of the quasi-Newton method to converge and minimize the
loss. This required computing the PageRanks of all the nodes in all the graphs
(100 of them) 123 times, along with the partial derivatives of each of the 8
parameters 123 times. On average, each PageRank computation took 13.2 steps of
power-iteration before converging, while each partial derivative computation
took 6.3 iterations. Each iteration for PageRank or its derivative takes
$O(|E|)$. Overall, the parameter estimation on Facebook network took 96
minutes. By contrast, increasing the number of edge types to 6 (which gives
best performance) required learning 48 weights, and increased the training time
to 13 hours on the Facebook dataset.



\section{Conclusion}
\label{sec:conclusion}
We have proposed Supervised Random Walks, a new learning algorithm for link
prediction and link recommendation.
By utilizing node and edge attribute data our method guides the random walks
towards the desired target nodes.
Experiments on Facebook and co-authorship networks demonstrate good
generalization and overall performance of Supervised Random Walks.
The resulting predictions show large improvements over Random Walks with
Restarts and compare favorably to supervised machine learning techniques that
require tedious feature extraction and generation.
In contrast, our approach requires no network feature generation and in a
principled way combines rich node and edge features with the structure of the
network to make reliable predictions.

Supervised Random Walks are not limited to link prediction, and can be applied
to many other problems that require learning to rank nodes in a graph, like
recommendations, anomaly detection, missing link, and expertise search and
ranking.

\xhdr{Acknowledgements} We thank Soumen Chakrabarti for discussion. Research
was in-part supported by NSF
CNS-1010921,  
NSF IIS-1016909,    
AFRL FA8650-10-C-7058, 
Albert Yu \& Mary Bechmann Foundation, IBM,
Lightspeed, Microsoft and Yahoo.

\bibliographystyle{abbrv}


\end{document}